\documentclass[numsec,webpdf,contemporary,large]{oup-authoring-template}%

\onecolumn 

\graphicspath{{Fig/}}

\theoremstyle{thmstyleone}%
\newtheorem{theorem}{Theorem}
\newtheorem{assumption}{Assumption}								   
\theoremstyle{thmstyletwo}%
\theoremstyle{thmstylethree}%

\usepackage{xcolor}
\usepackage[doublespacing]{setspace}
\usepackage[fontsize=12pt]{fontsize}
\usepackage{booktabs}
\newcommand{\blind}{0}
\newcommand{\independent}{\perp\!\!\!\perp}

\usepackage{hyperref}			
\usepackage{amsmath}
\usepackage{bbm}

\begin{document}

\journaltitle{Journals of the Royal Statistical Society}
\DOI{DOI HERE}
\copyrightyear{XXXX}
\pubyear{XXXX}
\access{Advance Access Publication Date: Day Month Year}
\appnotes{Paper}

\firstpage{1}


\title[Two-phase treatment with noncompliance]{Two-phase treatment with noncompliance: identifying the cumulative average treatment effect via multisite instrumental variables}

\author[1,$\ast$]{Guanglei Hong\ORCID{0000-0002-8254-4655}}
\author[2]{Xu Qin\ORCID{0000-0001-6160-1545}}
\author[3]{Zhengyan Xu\ORCID{0009-0003-3355-1482}}
\author[4,5]{Fan Yang\ORCID{0000-0003-3671-4745}}

\authormark{Hong et al.}

\address[1]{\orgdiv{Department of Comparative Human Development}, \orgname{University of Chicago}, \orgaddress{\street{Chicago}, \state{IL}, \country{USA}}}
\address[2]{\orgdiv{Department of Health and Human Development}, \orgname{University of Pittsburgh}, \orgaddress{\street{Philadelphia}, \state{PA}, \country{USA}}}
\address[3]{\orgdiv{Department of Economics}, \orgname{University of Pennsylvania}, \orgaddress{\street{Philadelphia}, \state{PA}, \country{USA}}}
\address[4]{\orgdiv{Yau Mathematical Sciences Center}, \orgname{Tsinghua University}, \orgaddress{\state{Beijing}, \country{China}}}
\address[5]{\orgname{Yanqi Lake Beijing Institute of Mathematical Sciences and Applications}, \orgaddress{\state{Beijing}, \country{China}}}

\corresp[$\ast$]{Address for correspondence. Guanglei Hong, Department of Comparative Human Development, University of Chicago, Chicago, IL 60637, USA. \href{Email:ghong@uchicago.edu}{ghong@uchicago.edu}}




\abstract{When evaluating a two-phase intervention, the cumulative average treatment effect (ATE) is often the primary causal estimand of interest. However, some individuals who do not respond well to the Phase I treatment may subsequently display noncompliant behaviours. At the same time, exposure to the Phase I treatment is expected to directly influence an individual's potential outcomes, thereby violating the exclusion restriction. Building on an instrumental variable (IV) strategy for multisite trials, we clarify the conditions under which the cumulative ATE of a two-phase treatment can be identified by employing the random assignment of the Phase I treatment as the instrument. Our strategy relaxes both the conventional exclusion restriction and sequential ignorability assumptions. We assess the performance of the new strategy through simulation studies. Additionally, we reanalyze data from the Tennessee class size study, in which students and teachers were randomly assigned to either small or regular class types in kindergarten (Phase I) with noncompliance emerging in Grade 1 (Phase II). Applying our new strategy, we estimate the cumulative ATE of receiving two consecutive years of instruction in a small versus regular class.}


\maketitle

\section{Introduction}
\subsection{Relevance}
A well-designed intervention often consists of more than one phase. This is because, to prevent fadeouts, the benefit of a treatment in an earlier phase may need to be reinforced through follow-up treatments in the later phases \citep{CunhaHeckman2007}. This rationale underpins many sustained treatment strategies in fields such as education, job training, counseling, and medicine for promoting long-term human development and well-being. In evaluating such programs, a research quantity of key scientific interest is the cumulative average treatment effect (ATE) of an intended experimental treatment sequence compared to a control sequence for a target population.

Authorized by the Tennessee state legislature in the 1980s, the Student Teacher Achievement Ratio study (Project STAR) was a state-wide multisite randomized trial designed to evaluate the multi-year cumulative impact of class size reduction on student achievement. Upon kindergarten entry, students within each participating school were randomly assigned to either a small class of 13-17 students (i.e., the experimental group) or a regular class of 22-25 students (i.e., the control group). Students were expected to remain in the class size type initially assigned throughout the study years. At the beginning of each year, teachers within each school were also randomly assigned by the project staff to teach a small class or a regular class \citep{FinnAchilles1990, MostellerLightSachs1995}. In this two-year study (kindergarten followed by Grade 1), each year represents a separate phase as students are assessed at the end of a year and are expected to possibly join a different class and meet a different teacher in the coming year. Our application study examines the cumulative ATE of receiving two years of instruction in a small class compared to a regular class in both kindergarten and Grade 1 in student achievement at the end of Grade 1.

\subsection{Methodological Challenges}
Despite unconditional randomization and full compliance in Phase I, several complications typically arise in multi-phase interventions.

\textit{Noncompliance following the Phase I treatment.} Noncompliance often becomes inevitable when an intervention program involves successive phases. In Project STAR, the annual progression in schooling naturally provides an occasion, at the end of each school year, for the school to move some students from one treatment condition to another, even though such decisions might be contrary to the intended research design. Indeed, some students changed their class size types after the kindergarten year \citep{Hanushek1999}. Identifying the cumulative ATE is challenging in the presence of noncompliance due to preexisting measured and unmeasured differences between those who complied with the initial treatment assignment and those who did not.

\textit{Posttreatment confounding.} If an individual does not respond well to a Phase I treatment, it may seem sensible to adjust their treatment in the subsequent phase when possible. In the context of class size interventions, for example, reasons for changing a student's class size type in Grade 1 are likely tied to the student's performance in kindergarten. In this case, kindergarten performance is a post-Phase 1 treatment (but pre-Phase II treatment) covariate that may confound the causal relationship between the Grade 1 treatment receipt and the outcome. Importantly, regression-based adjustments for such posttreatment confounding can introduce bias when assessing the effect of the kindergarten (Phase I) treatment on the outcome \citep{Rosenbaum1984}.

\textit{Non-additive treatment effects.} In some cases, continuing the same treatment into a later phase may not only sustain but also amplify the impact of the Phase I treatment. As a result, the benefit of receiving the treatment in two consecutive phases may exceed the sum of the benefits of receiving the treatment in Phase I alone and Phase II alone. In some other cases, a second phase of the same treatment may offer little additional value, such that the combined benefit of receiving two phases of the treatment is less than the sum of the single-phase treatment effects \citep{HongRaudenbush2008}. For these reasons, past studies that examined the impact of class size reduction in either kindergarten or Grade 1 in isolation have not addressed the key question about the cumulative ATE of receiving the treatment across both years \citep{Krueger1999}.

\subsection{Constraints of Existing Causal Inference Strategies}
These complications pose methodological challenges to evaluations of multi-phase intervention programs even when the Phase I treatment is randomized. Existing causal inference strategies show limitations in overcoming these challenges.

\textit{A naive intention-to-treatment (ITT) analysis} compares the average observed outcomes between the experimental and control groups based on initial treatment assignment. However, in the presence of noncompliance, the ITT effect typically underestimates the cumulative ATE of the experimental treatment sequence relative to the control sequence.

\textit{Inverse-probability-of-treatment weighting (IPTW)} is commonly used to address posttreatment confounding when evaluating the cumulative effects of time-varying treatments \citep{HongRaudenbush2008, RobinsHernanBrumback2000} or per-protocol effects in the presence of nonadherence \citep{HernanRobins2017PerProtocol, SmithCoffmanHudgens2021}. However, identification using IPTW relies on the sequential ignorability assumption that rarely holds in practice, as noncompliant behavior is often related to unobserved confounders, even after adjusting for observed ones. In prior analyses of the Project STAR data, researchers have employed statistical adjustment for observed pretreatment covariates but failed to account for both posttreatment confounding and unobserved pretreatment confounding \citep{NyeHedgesKonstantopoulos2000}.

When applying a \textit{standard instrumental variable (IV) strategy} to address noncompliance, the analyst may use the initial random treatment assignment as the instrument for the non-random treatment received in a later phase. The IV estimand is defined as the ratio of the ITT effect of the instrument on the outcome to the ITT effect on the treatment received \citep{Bloom1984}. This strategy requires the exclusion restriction, which is unlikely to hold in multi-phase interventions. This is because the instrument is expected to have a direct effect on the outcome through actual exposure to the Phase I treatment, regardless of compliance in Phase II. Therefore, the standard IV method is not appropriate for identifying the cumulative ATE in such settings \citep{SwansonLabrecqueHernan2018}. Previous studies using Project STAR data have applied the IV method to reduce selection bias associated with noncompliance in year-by-year analyses. However, these approaches do not address the cumulative ATE, as the yearly effects may not be additive \citep{Krueger1999, ShinRaudenbush2011}.

Researchers have sought to estimate the average effect of the Phase II treatment on the outcome in the target population by extrapolating from the complier population. This is done by first estimating \textit{compliance scores} as functions of observed pretreatment covariates that predict both compliant behavior and heterogeneity in treatment effects, and then estimating the average effect of the Phase II treatment within levels of compliance scores \citep{Angrist2010NBER, AronowCarnegie2013, EsterlingNebloLazer2011, WangRobinsRichardson2017}. These strategies nonetheless rely on a form of the strong ignorability assumption that is often unrealistic. Specifically, unobserved pretreatment confounders as well as observed and unobserved time-varying confounders may not be captured by the estimated compliance scores, which undermines the validity of the identification.

The \textit{multiple-site multiple-mediator IV (MSMM-IV) strategy} \citep{BloomUntermanZhuReardon2020, DuncanMorrisRodrigues2011, KlingLiebmanKatz2007, NomiRaudenbushSmith2021, RaudenbushReardonNomi2012, ReardonRaudenbush2013}, developed to identify the causal effects of multiple parallel mediators in a multisite trial, similarly requires the exclusion restriction and overlooks non-additive cumulative treatment effects. This strategy generally does not allow for time-varying confounders of the mediator-outcome relationships that may affect the outcome both directly and indirectly (see \citet{ReardonRaudenbush2013} for a discussion of a possible relaxation of this constraint).

\subsection{Proposed Multisite Two-Phase Treatment IV Strategy}
Capitalizing on a multisite randomized design that supplies the initial random treatment assignment as an instrument per site, we propose a multisite two-phase treatment IV (MS2T-IV) strategy. Our goal is to identify and estimate the cumulative ATE by extending the existing MSMM-IV framework to accommodate two-phase treatment sequences with noncompliance following the Phase I treatment.

This methodological development overcomes some major limitations of existing strategies. Unlike IPTW and compliance score-based methods that are vulnerable to bias from unmeasured confounders, the MS2T-IV strategy is designed to eliminate unmeasured pretreatment confounding, thereby relaxing the overly strong assumption of sequential ignorability. Moreover, unlike standard IV strategies, MS2T-IV adjusts for observed posttreatment confounders and allows for non-additive treatment effects as well as direct effects of Phase I treatment exposure on the outcome, thereby relaxing the exclusion restriction.

When using the initial random treatment assignment as an instrument to address complications arising from noncompliance, the well-known principal stratification framework \citep{FrangakisRubin2002} assumes deterministic noncompliance. That is, each individual is classified as either a complier or a noncomplier, based on the notion that compliance is pre-determined and fixed. Under this framework, the IV estimand identifies the Complier Average Treatment Effect (CATE), also known as the Local Average Treatment Effect (LATE) \citep{AngristImbensRubin1996}. 

However, as \citet{SmallTanEtAl2017} argue, this deterministic view may not align with reality, especially when compliant behavior is influenced by stochastic events. In such cases, the IV estimand instead identifies a Weighted Average Treatment Effect (WATE), a quantity that often does not directly correspond to the target population of the study. In our application, we similarly conceptualize noncompliance as stochastic. For the primary target population in Project STAR, due to unforeseen opportunities and constraints, whether a student would change class type in Grade 1 is probable rather than pre-determined. Therefore, we argue that the concepts of CATE or LATE may not be applicable while the concept of cumulative ATE of attending a small versus regular class over consecutive years is relevant.

Our proposed approach is specific to multisite randomized trials. But it may also be applicable to settings analogous to two-phase treatment designs. For example, \citet{HeckmanSmithTaber1998} demonstrated that intervention programs may affect not only experimental participants who receive the full treatment but also those who receive a partial ``dose" before dropping out. When the effect of partial treatment is nonzero, the IV estimator fails to identify the average treatment effect for the fully treated due to the violation of the exclusion restriction. In such a case, Heckman and colleagues argued that an economically relevant causal parameter is the average treatment effect that would result if the entire experimental group would receive the full treatment. This parameter aligns with our concept of cumulative ATE, with individuals receiving partial treatment prior to dropout analogous to those who switch from the experimental condition to the control condition after receiving a Phase-I treatment.

This article is organized as follows. Section \ref{sec2} derives the theoretical results for identifying the cumulative ATE and presents the steps of MS2T-IV estimation and statistical inference. Section \ref{sec3} assesses the performance of estimation and inference through simulations. Section \ref{sec4} applies the proposed strategy to the Project STAR data. Section \ref{sec5} concludes and discusses future research.

\section{Identification and Estimation of the Cumulative ATE via the MS2T-IV Strategy}\label{sec2}
\subsection{Notation and the Causal Estimand}
Let $Z_{i k}$ be a binary indicator for the Phase I treatment that takes value $z=1$ if individual $i$ at site $k$ is assigned to the experimental condition and 0 if assigned to the control condition at the baseline. In Project STAR, due to full compliance in kindergarten, the treatment assigned is equivalent to the treatment received for each student in Phase I. Let $D_{i k}$ be a binary indicator for the Phase II treatment receipt that takes value $d=1$ if the individual receives the experimental treatment and 0 if receiving the control treatment. The individual's potential outcome by the end of Phase II can be represented as a function of the initial randomized treatment assignment $z$ or a function of the two-phase treatment sequence $z$ and $d$. It is typically denoted as $Y_{i k}(z)$ in the former case and $Y_{i k}(z, d)$ in the latter case for $z=0,1$ and $d=0,1$ under the stable unit treatment value assumption (SUTVA). In other words, we assume that there is a single version of each treatment and that there is no interference among individuals between or within sites. Similarly, we may use $V_{i k}(z)$ to denote the individual's intermediate potential outcome by the end of Phase I and use $D_{i k}(z)$ to denote the individual's Phase-II treatment receipt each as a function of the individual's Phase-I treatment assignment. We use $\mathrm{X}_{i k}$ and $\mathrm{U}_{i k}$ to denote the individual's observed and unobserved vectors of baseline covariates, respectively. We use $S_{i}=k$ to indicate that individual $i$ is a member of site $k$.

Corresponding to a research question that initially motivated Project STAR, the causal effect of attending a small class versus a regular class in both kindergarten and Grade 1 for student $i$ in school $k$ is $Y_{i k}(1,1)-Y_{i k}(0,0)$. The cumulative ATE of the two-phase treatment, as defined below, represents the average of site-specific treatment effects for a population of sites where the inner expectation is taken over the population of individuals within site $k$ and the outer expectation is taken over all the sites:
$$
\delta_{A T E}=E\left(\Delta_{A T E_{k}}\right)=E\left[E\left\{Y_{i k}(1,1)-Y_{i k}(0,0)\right\}\right] .
$$

\subsection{Theoretical Model}
\begin{figure}[t]
\begin{center}
\includegraphics[scale=0.25]{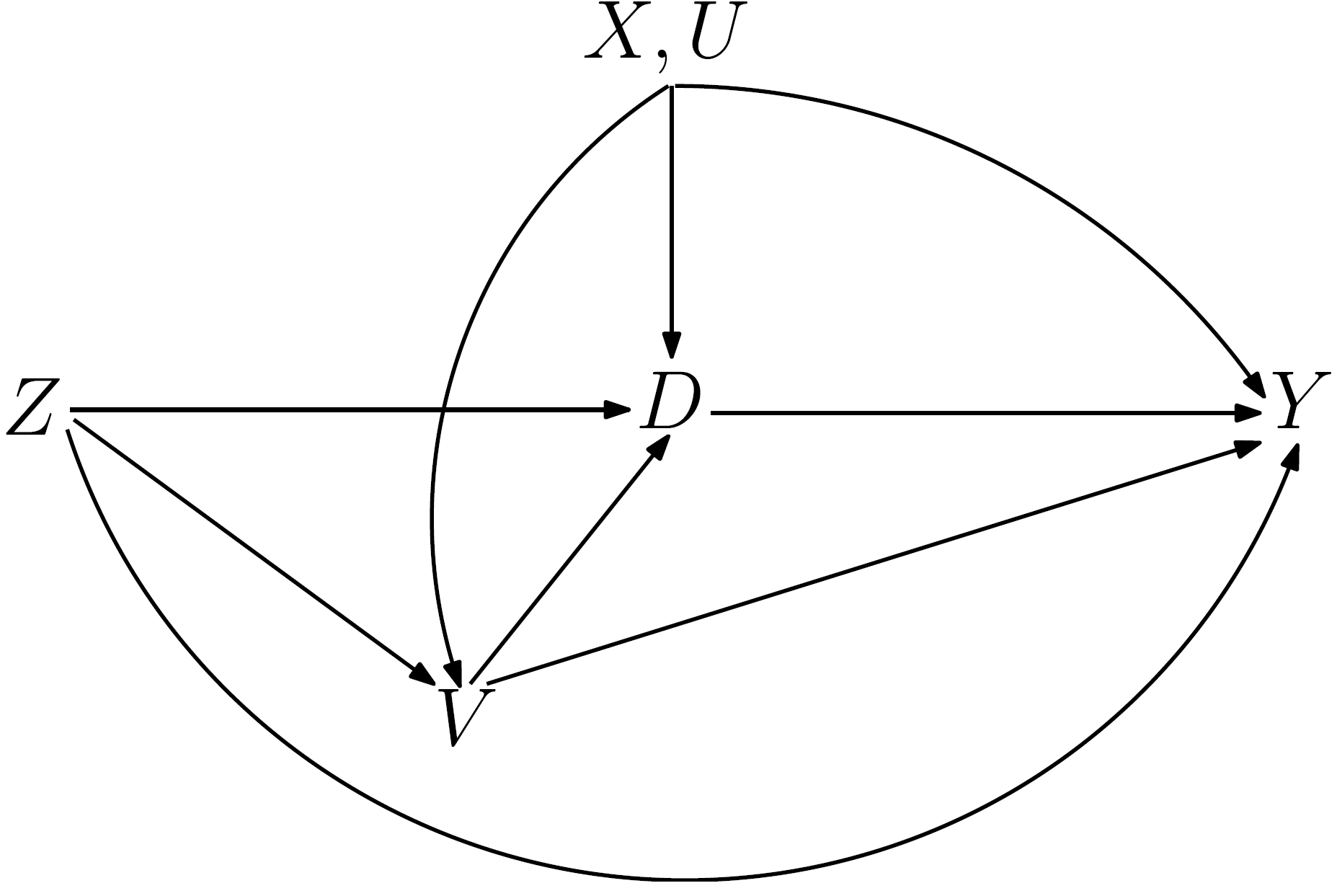}
\caption{Directed Acyclic Graph for Two-Phase Treatment Sequences with Noncompliance. $Z$ and $D$ respectively denote the Phase I treatment and Phase II treatment receipt. $V$ and $Y$ respectively stand for the outcome by the end of Phase I and that by the end of Phase II. $X$ and $U$ denote the observed and unobserved vectors of baseline covariates, respectively.}\label{fig1}
\end{center}
\vspace{-3em}
\end{figure}

Figure \ref{fig1} illustrates the conceptual relationships between $X, U, Z, V, D$, and $Y$. We have omitted the subscripts in the figure for brevity. For individual $i$ in site $k$, $V_{i k}(z)$ for $z=0,1$ is a function of the individual's fixed pretreatment covariates $\mathrm{X}_{i k}$ and $\mathrm{U}_{i k}$ and site membership; and $Y_{i k}(z, d)= Y_{i k}\left(z, V_{i k}(z), d\right)$ for $z, d=0,1$ is a function of $\mathrm{X}_{i k}, \mathrm{U}_{i k}, V_{i k}(\mathrm{z})$ and site membership. Importantly, $D_{i k}(z)$ for $z=0,1$ is subject to the selection associated with $\mathrm{X}_{i k}, \mathrm{U}_{i k}$, and $V_{i k}(z)$ within a site. Because $D_{i k}(z)$ and $Y_{i k}(z, d)$ are both functions of $V_{i k}(z)$, as shown clearly in Figure \ref{fig1}, $V_{i k}(z)$ is a posttreatment confounder of the causal relationship between $D_{i k}(z)$ and $Y_{i k}(z, d)$ under treatment condition $z$.

Let $\beta_{0 i k}=D_{ik}(0)$ denote an individual's Phase-II treatment after being assigned to the control condition in Phase I; let $\beta_{1 i k}=D_{ik}(1)-D_{ik}(0)$ denote the individual-specific ITT effect of $Z_{i k}$ on $D_{i k}$; and let $\beta_{2 i k}=D_{ik}(1)$ denote the individual's Phase-II treatment after being assigned to the experimental condition in Phase I. Here $\beta_{0 i k}$, $\beta_{1 i k}$, and $\beta_{2 i k}$ are each a function of $\mathrm{X}_{i k}$ and $\mathrm{U}_{i k}$. We then define $\beta_{0 k}=E\left\{\beta_{0 i k}\mid S_i=k\right\}$ as the site-specific noncompliance rate in Phase II following the assignment to the control condition in Phase I, $\beta_{1 k}=E\left\{\beta_{1 i k}\mid S_i=k\right\}$ as the site-specific ITT effect of $Z$ on $D$, and $\beta_{2 k}=E\left\{\beta_{2 i k}\mid S_i=k\right\}$ as the site-specific compliance rate in Phase II following the assignment to the experimental condition in Phase I. If site $k$ displays perfect compliance, then we have that $\beta_{0 k}=0$ and $\beta_{1k}=\beta_{2k}=1$.

In addition, let $\alpha_{0 i k} = V_{i k}(0)$ denote an individual's Phase-I outcome after being assigned to the control condition in Phase I, which is explained by $\mathrm{X}_{i k}$ and $\mathrm{U}_{i k}$. Let $\alpha_{1 i k} = V_{i k}(1)-V_{i k}(0)$, which is the individual-specific ITT effect of $Z_{i k}$ on $V_{i k}$ and is also explained by $\mathrm{X}_{i k}$ and $\mathrm{U}_{i k}$. $V$ can be either continuous or discrete. When $V$ is multi-categorical or multivariate, $\alpha_{0 i k}$ and $\alpha_{1 i k}$ become vectors.

We now specify the simplest form of the theoretical model (i.e., the data generation) for $Y_{ik}\left(z,d\right)$ as the following for $z,d=0, 1$. One may expand this model by including site mean-centered $X_{ik}$ as additional predictors. The theoretical model can be extended to the case in which $V_{ik}\left(z\right)$ is multivariate.
\vspace{-1em}
\begin{align*}
& Y_{i k}(z, d)=\theta_{Y 0 i k}+\theta_{V i k} V_{i k}(z)+\gamma_{1 i k} z+\gamma_{2 i k} d+\gamma_{3 i k} z d ; \tag{1}\label{eq1}
\end{align*}
\vspace{-3em}

In the theoretical model for $Y_{i k}(z, d)$, the intercept $\theta_{Y 0 i k}$ captures the individual-specific element of $Y_{i k}(0,0)$ explained by $X_{i k}$ and $U_{i k}$. We use $\gamma_{1 i k}, \gamma_{2 i k}$, and $\gamma_{3 i k}$ to denote the components of the cumulative ATE for the individual attributable to the individual's responses to the Phase I treatment, the Phase II treatment, and their interaction, respectively, when $V_{i k}(z)$ is given. Here $\gamma_{1 i k}, \gamma_{2 i k}$, and $\gamma_{3 i k}$ are each a function of $\mathrm{X}_{i k}$ and $\mathrm{U}_{i k}$. The remaining part of the cumulative ATE for the individual is captured by $\theta_{V i k}\left\{V_{i k}(1)-V_{i k}(0)\right\}=\theta_{V i k} \alpha_{1 i k}$, which is the indirect effect of $Z_{i k}$ on $Y_{i k}$ transmitted through the individual-specific ITT effect of $Z_{i k}$ on $V_{i k}$ but not through $D_{i k}$. This model implies two assumptions. (a) By definition, $Y_{i k}(z, d)=Y_{i k}\left(z, V_{i k}(z), d\right)$ for $z, d=0,1$, which implies that when $z, d, \mathrm{X}_{i k}, \mathrm{U}_{i k}$, and $V_{i k}(z)$ are given, $V_{i k}\left(z^{\prime}\right)$ does not provide additional information about $Y_{i k}(z, d)$. (b) When $z$ and $d$ are given, the association between $V_{i k}$ and $Y_{i k}$, which is represented by $\theta_{V i k}$, is assumed to be linear and not depend on $z$ or $d$. Although this theoretical model (\ref{eq1}) is assumed to be linear and partially additive, analysts may explore alternative model specifications. For instance, to accommodate potential nonlinear relationships between $V$ and $Y$, more flexible functional forms can be employed. 

Aggregating the individual-specific parameters to the site level and using $h$ for numbering, we have that $\alpha_{h k}= E\left(\alpha_{h i k} \mid S_{i}=k\right)$ for $h=0,1 ; \theta_{Y 0 k}=E\left(\theta_{Y 0 i k} \mid S_{i}=k\right) ;$ and $\gamma_{h k}=E\left(\gamma_{h i k} \mid S_{i}=k\right)$ for $h=1,2,3$.

Table \ref{tab:glossary} provides a glossary of these individual-specific and site-specific parameters.

Finally, we define $\theta_{V}=E\left(\theta_{V k}\right)$ and $\gamma_{h}=E\left(\gamma_{h k}\right)$ for $h=1,2,3$.

\begin{table}[ht]
\centering
\caption{Glossary of Individual-Specific and Site-Specific Parameters}\label{tab:glossary}
\renewcommand{\arraystretch}{1.5} 
\begin{tabular}{@{}l p{0.87\linewidth}@{}}
\hline
Notation & Definition \\
\hline
$\alpha_{0ik}$ & $V_{ik}(0)$ that is individual-specific \\
$\alpha_{1ik}$ & $V_{ik}(1)-V_{ik}(0)$, individual-specific ITT effect of $Z_{ik}$ on $V_{ik}$ \\
$\beta_{0ik}$ & $D_{ik}(0)$ that is individual-specific \\
$\beta_{1ik}$ & $D_{ik}(1)-D_{ik}(0)$, individual-specific ITT effect of $Z_{ik}$ on $D_{ik}$ \\
$\beta_{2ik}$ & $D_{ik}(1)$ that is individual-specific \\
$\gamma_{1ik}$ & $\left\{Y_{ik}(1,0)-Y_{ik}(0,0)\right\}-\theta_{Vk}\left\{V_{ik}(1)-V_{ik}(0)\right\}$, individual-specific direct effect of $Z_{ik}$ on $Y_{ik}$ that does not operate through $V_{ik}$ should the individual receive the control condition in the Phase II treatment \\
$\gamma_{2ik}$ & $Y_{ik}(0,1)-Y_{ik}(0,0)$, individual-specific direct effect of $D_{ik}$ on $Y_{ik}$ should the individual be assigned to the control condition in the Phase I treatment \\
$\gamma_{3ik}$ & $\left\{Y_{ik}(1,1)-Y_{ik}(1,0)\right\}-\left\{Y_{ik}(0,1)-Y_{ik}(0,0)\right\}$, individual-specific interaction effect of $Z_{ik}$ and $D_{ik}$ on $Y_{ik}$ \\
$\theta_{Vik}$ & individual-specific linear association between $V_{ik}(z)$ and $Y_{ik}(z,d)$\\
$\alpha_{1k}$ & $E\left(\alpha_{1ik}\mid S_i=k\right)$, site-specific ITT effect of $Z_{ik}$ on $V_{ik}$ \\
$\beta_{1k}$ & $E\left\{D_{ik}(1)-D_{ik}(0)\mid S_i=k\right\}$, site-specific ITT effect of $Z_{ik}$ on $D_{ik}$ \\
$\beta_{2k}$ & $E\left\{D_{ik}(1)\mid S_i=k\right\}$, site-specific mean of $D_{ik}(1)$ \\
$\gamma_{1k}$ & $E\left(\gamma_{1ik}\mid S_i=k\right)$, site-specific mean of $\gamma_{1ik}$ \\
$\gamma_{2k}$ & $E\left(\gamma_{2ik}\mid S_i=k\right)$, site-specific mean of $\gamma_{2ik}$ \\
$\gamma_{3k}$ & $E\left(\gamma_{3ik}\mid S_i=k\right)$, site-specific mean of $\gamma_{3ik}$ \\
$\theta_{Vk}$ & $E\left(\theta_{vik}\mid S_i=k\right)$, site-specific mean of $\theta_{Vik}$ \\ 
\hline
\end{tabular}
\vspace{-2em}
\end{table}

\subsection{Identification Assumptions}
When the theoretical model is aligned with the data generation process, Assumptions \ref{ass1}, \ref{ass2}, and \ref{ass3} as stated below are sufficient for identifying the cumulative ATE.

\begin{assumption}\label{ass1}
Within-site ignorable Phase I treatment assignment. The Phase I
treatment assignment $Z$ is independent of all the potential outcomes within each site:
$
\left\{V_{i k}(0), D_{i k}(0), Y_{i k}(0), V_{i k}(1), D_{i k}(1), Y_{i k}(1)\right\} \independent Z_{i k} \mid S_{i}=k.
$
\end{assumption}
This assumption is guaranteed by the multisite randomized design.
\begin{assumption}\label{ass2}
Within-site zero covariance assumption. $\mathrm{Cov}\left\{\gamma_{2 i k}, \beta_{1 i k}\right\} =0$,  $\mathrm{Cov}\left\{\gamma_{3 i k}, \beta_{2 i k}\right\}=0$, and $\mathrm{Cov}\left\{\theta_{v i k}, \alpha_{1 i k}\right\} =0$.
\end{assumption}
Assumption \ref{ass2} rules out individual-level pretreatment covariates that predict not just the individual-specific ITT effect of $Z$ on $D$ but also the individual-specific causal effect of $D$ on $Y$ under assignment to the control condition in Phase I. It likewise rules out covariates that  predict not only the individual-specific compliance in Phase II after being assigned to the experimental condition in Phase I but also the individual-specific interaction effect of $Z$ and $D$ on $Y$. Moreover, it rules out covariates that predict not only the individual-specific ITT effect of $Z$ on $V$ but also the individual-specific linear association between $V$ and $Y$. Intuitively speaking, this assumption requires that the Phase I process is uncorrelated with the Phase II process within a site.

Violations of Assumption \ref{ass2} are conceivable. For example, students who behave well due to attending a small kindergarten class are perhaps less likely to change to a regular class in Grade 1; moreover, such students may benefit more from attending a small class in Grade 1, in which case $\gamma_{3 i k}$ would co-vary with $\beta_{2 i k}$. In another example, if students who are more likely to comply with the treatment assignment are to benefit more from attending a small class in Grade 1 , then $\gamma_{2 i k}$ would co-vary with $\beta_{1 i k}$.
\vspace{-0.8em}
\begin{assumption}\label{ass3}
Between-site independence assumption. $\alpha_{1 k}$, $\beta_{1 k}$, and $\beta_{2 k}$ are jointly independent of $\theta_{V k}$, $\gamma_{2 k}$, and $\gamma_{3 k}$; furthermore, $\alpha_{1 k}$, $\beta_{1 k}$, and $\beta_{2 k}$ are each independent of $\gamma_{1 k}$.
\end{assumption}
\vspace{-0.8em}
Assumption \ref{ass3} implies the absence of site-level characteristics that predict not only the site-specific ITT effect of $Z$ on $V$, the site-specific ITT effect of $Z$ on $D$, or the site-specific compliance rate among individuals assigned to the experimental condition in Phase I, but also the site-specific association between $V$ and $Y$, or any other element of the site-specific cumulative ATE including $\gamma_{1 k}$, $\gamma_{2 k}$, and $\gamma_{3 k}$. The joint independence further implies that, when $\alpha_{1 i k}$, $\beta_{1 i k}$, or $\beta_{2 i k}$ covary within a site, such covariance is independent of $\theta_{V k}$, $\gamma_{2 k}$, and $\gamma_{3 k}$; the former is also independent of the within-site covariance between $\theta_{V i k}$, $\gamma_{2 i k}$, or $\gamma_{3 i k}$; similarly, the within-site covariance of the latter type is independent of $\alpha_{1 k}$, $\beta_{1 k}$, and $\beta_{2 k}$. Assumption \ref{ass3} can be violated. For example, if schools with higher compliance rates in Grade 1 tend to derive a greater direct benefit of the kindergarten treatment on Grade 1 achievement, then $\beta_{1 k}$ and $\gamma_{1 k}$ are not independent.

\vspace{-3em}
\begin{theorem}\label{theorem}
Under the theoretical model (\ref{eq1}) and when Assumptions \ref{ass1}, \ref{ass2}, and \ref{ass3} hold, the cumulative ATE is the sum of four components as shown in equation (\ref{eq6}):
\begin{align*}
\delta_{A T E} & =E\left[E\left\{Y_{i k}(1,1)-Y_{i k}(0,0) \mid S_{i}=k\right\}\right] \\
& =E\left[E\left[\left\{\theta_{Y 0 i k}+\theta_{V i k} V_{i k}(1)+\gamma_{1 i k}+\gamma_{2 i k}+\gamma_{3 i k}\right\}-\left\{\theta_{Y 0 i k}+\theta_{V i k} V_{i k}(0)\right\} \mid S_{i}=k\right]\right] \\
& =E\left(\gamma_{1 k}+\gamma_{2 k}+\gamma_{3 k}+\theta_{V k} \alpha_{1 k}\right) \\
& =\gamma_{1}+\gamma_{2}+\gamma_{3}+\theta_{V} \alpha_{1} . \tag{2}\label{eq2}
\end{align*}
\end{theorem}
\vspace{-3em}
\subsection{Reduced-Form Models for the ITT Effects}
This section shows that, once we identify the site-specific ITT effects of $Z$ on $V, D$, and $Y$, we can subsequently identify the elements in the cumulative ATE. Below we derive the reduced-form models for $V_{i k}, D_{i k}$ and $Y_{i k}$ within each site.

\textit{Reduced-form model for V.} Because $E\left\{V_{i k}(z) \mid S_{i}=k\right\}=\alpha_{0 k}+\alpha_{1 k} z$, the site-specific ITT effect of $Z$ on $V$ is $E\left\{V_{i k}(1)-V_{i k}(0) \mid S_{i}=k\right\}=\alpha_{1 k}$. Given that $Z_{ik}$ is randomized, the analytic model for obtaining the ITT effect of $Z$ on $V$ in site $k$ is given by
\begin{equation*}
V_{i k}=\alpha_{0 k}+\alpha_{1 k} Z_{i k}+e_{V i k}, \tag{3}\label{eq3}
\end{equation*}
where $e_{V i k}$ is an error term that, under Assumption \ref{ass1}, has mean zero and is uncorrelated with $Z_{i k}$ within site $k$. 

\textit{Reduced-form model for $D$.} Because $E\left\{D_{i k}(z) \mid S_{i}=k\right\}=\beta_{0 k}+\beta_{1 k} z$ where $\beta_{0 k}=\beta_{2k}-\beta_{1k}$, the site-specific ITT effect of $Z$ on $D$ is $E\left\{D_{i k}(1)-D_{i k}(0) \mid S_{i}=k\right\}=\beta_{1 k}$. Given that $Z_{ik}$ is randomized, the analytic model for obtaining the site-specific ITT effect of $Z$ on $D$ in site $k$ is given by
\begin{equation*}
D_{i k}=\beta_{0 k}+\beta_{1 k} Z_{i k}+e_{D i k}, \tag{4}\label{eq4}
\end{equation*}
where $e_{D i k}$ is an error term that, under Assumption \ref{ass1}, has mean zero and is uncorrelated with $Z_{i k}$ within site $k$.

\textit{Reduced-form model for $Y$}. As we show in section S1.1 of the supplementary material, under Assumptions \ref{ass1} and \ref{ass2}, the site-specific ITT effect of $Z$ on $Y$ is
$$
E\left\{Y_{i k}(1)-Y_{i k}(0) \mid S_{i}=k\right\}=\theta_{V k} \alpha_{1 k}+\gamma_{1 k}+\gamma_{2 k} \beta_{1 k}+\gamma_{3 k} \beta_{2 k},
$$
where $\beta_{2 k}=\beta_{0 k}+\beta_{1 k}$. 
Given that $Z_{ik}$ is randomized, the analytic model for obtaining the ITT effect of $Z$ on $Y$ in site $k$ is given by:
\begin{equation*}
Y_{i k}=\theta_{0 k}+\theta_{1 k} Z_{i k}+e_{Y i k}, \tag{5}\label{eq5}
\end{equation*}
where $\theta_{0 k}=E\left\{Y_{i k}(0) \mid S_{i}=k\right\}; \theta_{1 k}=\gamma_{1 k}+\gamma_{2 k} \beta_{1 k}+\gamma_{3 k} \beta_{2 k}+\theta_{V k} \alpha_{1 k}$. The error term $e_{Y i k}$, under Assumption 1, has mean zero and is uncorrelated with $Z_{i k}$ within site $k$.

\textit{Site-level population model.} We have derived that $\theta_{1 k}$ is a function of $\alpha_{1 k}, \beta_{1 k}$, and $\beta_{2 k}$ in a population of sites. Let
\begin{equation*}
\theta_{1 k}=\gamma_{1}+\gamma_{2} \beta_{1 k}+\gamma_{3} \beta_{2 k}+\theta_{V} \alpha_{1 k}+\epsilon_{k}, \tag{6}\label{eq6}
\end{equation*}
where $\epsilon_{k}=\left(\gamma_{1 k}-\gamma_{1}\right)+\left(\gamma_{2 k}-\gamma_{2}\right) \beta_{1 k}+\left(\gamma_{3 k}-\gamma_{3}\right) \beta_{2 k}+\left(\theta_{V k}-\theta_{V}\right) \alpha_{1 k}$. We show in section S1.2 of the supplementary material that, under Assumption \ref{ass3}, $\epsilon_{k}$ has mean zero and is uncorrelated with the predictors $\beta_{1 k}, \beta_{2 k}$, and $\alpha_{1 k}$ in model (\ref{eq6}). After identifying $\alpha_{1 k}, \beta_{1 k}$, and $\beta_{2 k}$ for each site, we can subsequently obtain the values of $\gamma_{1}, \gamma_{2}, \gamma_{3}$, and $\theta_{V}$, and thereby identifying $\delta_{A T E}$ per Equation (\ref{eq2}).

\vspace{-1em}
\subsection{The Distinction Between the Cumulative ATE and the ITT Effect on the Outcome}
The cumulative ATE differs from the ITT effect in a two-phase study with noncompliance following Phase I. Specifically, the cumulative ATE represents the causal effect of the intended experimental treatment sequence compared to the control condition sequence, whereas the ITT effect reflects the causal effect of the initial treatment assignment regardless of the actual treatment sequences. The difference between the ITT effect and the cumulative ATE can be expressed as 
\begin{align*}
\mathrm{E}(\theta_{1k})-\delta_{\mathrm{ATE}} &= (\gamma_1+\gamma_{2}\beta_{1}+\gamma_{3}\beta_{2}+\theta_{V}\alpha_{1}) - (\gamma_{1}+\gamma_{2}+\gamma_{3}+\theta_{V} \alpha_{1})\\& = \gamma_{2}(\beta_{1}-1)+\gamma_{3}(\beta_2-1).
\end{align*}
As expected, this difference vanishes under full compliance where $\beta_1=\beta_2=1$. The cumulative ATE also differs from the CATE or LATE, as it is defined for the entire targeted study population, while the CATE or LATE is specific to the subpopulation of compliers.

\vspace{-1em}

\subsection{Estimation and Statistical Inference}
Based on the above theoretical results, we propose a multisite IV analysis for estimating the cumulative ATE of a two-phase treatment sequence. The analysis involves two stages. In Stage 1, we estimate the ITT effects of $Z$ on $V, D$, and $Y$ within each site. In Stage 2, we obtain estimates of $\gamma_{1}, \gamma_{2}, \gamma_{3}$ and $\theta_{V}$ through a between-site analysis.

\textit{Stage 1 analysis.} Through a site-by-site OLS analysis of the sample data with $Z_{i k}$ as the predictor of $V_{i k}, D_{i k}$, and $Y_{i k}$ for individual $i$ in site $k$, we estimate the respective site-specific ITT effects ${\alpha}_{1 k}, {\beta}_{1 k}$, and ${\theta}_{1 k}$ as well as the intercept ${\beta}_{0 k}$ and then compute $\hat{\beta}_{2 k}=\hat{\beta}_{0 k}+\hat{\beta}_{1 k}$. These analytic models are consistent with the reduced-form models in equations (\ref{eq3}), (\ref{eq4}), and (\ref{eq5}).

\textit{Stage 2 analysis.} We then analyze via OLS a site-level regression model as specified in equation (\ref{eq6}), replacing $\theta_{1 k}, \beta_{1 k}, \beta_{2 k}$, and $\alpha_{1 k}$ with their sample analogues obtained from the Stage 1 analysis. Corresponding to equation (\ref{eq2}), the sample estimator of the cumulative ATE is

\begin{equation*}
\hat{\delta}_{A T E}=\hat{\gamma}_{1}+\hat{\gamma}_{2}+\hat{\gamma}_{3}+\hat{\theta}_{V} \hat{\alpha}_{1} . \tag{7}\label{eq7}
\end{equation*}

\textit{MS2T-IV with covariance adjustment.} Using the sample estimators $\hat{\alpha}_{1 k}, \hat{\beta}_{1 k}$, and $\hat{\beta}_{2 k}$ as predictors in the Stage 2 analysis may lead to finite-sample bias in the estimation of $\delta_{A T E}$. We recommend an additional adjustment for site mean-centered $X$ and their potential interactions with $Z$ in the Stage 1 analysis. This additional adjustment is expected to reduce such finite-sample bias, and increase estimation efficiency. Let $\overline{\mathrm{X}}_{. k}$ denote the site mean of $X$ in site $k$. Specifically, in the reduced-form models (\ref{eq3}), (\ref{eq4}), and (\ref{eq5}), let $e_{V i k}=\alpha_{\mathrm{x} k}^{\prime}\left(\mathrm{X}_{i k}-\overline{\mathrm{X}}_{. k}\right)+ \alpha_{\mathrm{x} z k}^{\prime}\left(\mathrm{X}_{i k}-\overline{\mathrm{X}}_{. k}\right) Z_{i k}+\varepsilon_{V i k}, e_{D i k}=\beta_{\mathrm{x} k}^{\prime}\left(\mathrm{X}_{i k}-\overline{\mathrm{X}}_{. k}\right)+\beta_{\mathrm{x} z k}^{\prime}\left(\mathrm{X}_{i k}-\mathrm{X}_{. k}\right) Z_{i k}+\varepsilon_{D i k}$, and $e_{Y i k}= \theta_{\mathrm{x} k}^{\prime}\left(\mathrm{X}_{i k}-\overline{\mathrm{X}}_{. k}\right)+\theta_{\mathrm{xz} k}^{\prime}\left(\mathrm{X}_{i k}-\overline{\mathrm{X}}_{. k}\right) Z_{i k}+\varepsilon_{Y i k}$, respectively. The analytic models in the Stage 1 analysis then become:
$$\begin{aligned}
& V_{i k}=\alpha_{0 k}+\alpha_{1 k} Z_{i k}+\alpha_{\mathrm{x} k}^{\prime}\left(\mathrm{X}_{i k}-\overline{\mathrm{X}}_{. k}\right)+\alpha_{\mathrm{x} z k}^{\prime}\left(\mathrm{X}_{i k}-\overline{\mathrm{X}}_{. k}\right) Z_{i k}+\varepsilon_{V i k} ; \\
& D_{i k}=\beta_{0 k}+\beta_{1 k} Z_{i k}+\beta_{\mathrm{x} k}^{\prime}\left(\mathrm{X}_{i k}-\overline{\mathrm{X}}_{. k}\right)+\beta_{\mathrm{x} z k}^{\prime}\left(\mathrm{X}_{i k}-\mathrm{X}_{. k}\right) Z_{i k}+\varepsilon_{D i k} ; \\
& Y_{i k}=\theta_{0 k}+\theta_{1 k} Z_{i k}+\theta_{\mathrm{x} k}^{\prime}\left(\mathrm{X}_{i k}-\overline{\mathrm{X}}_{. k}\right)+\theta_{\mathrm{x} z k}^{\prime}\left(\mathrm{X}_{i k}-\overline{\mathrm{X}}_{. k}\right) Z_{i k}+\varepsilon_{Y i k} .
\end{aligned}
$$
Because $\mathrm{X}_{i k}$ is site mean-centered, the new error terms $\varepsilon_{V i k}, \varepsilon_{D i k}$, and $\varepsilon_{Y i k}$ each have a mean of zero and are uncorrelated with $Z_{i k}$ under Assumption \ref{ass1}. The site-specific ITT effects $\alpha_{1 k}, \beta_{1 k}$, and $\theta_{1 k}$ remain unchanged; yet the efficiency of their sample estimators will improve due to the reduction in error variance. At Stage 2, we analyze the same site-level model as specified in equation (\ref{eq6}) and obtain the sample estimator of the cumulative ATE in the same form as that in equation (\ref{eq7}).

\textit{Improper statistical inference.} 
The variance of the estimator of $\delta_{\text {ATE }}$ is 
$$
\begin{aligned}
\operatorname{Var}\left(\hat{\delta}_{A T E}\right)= & \operatorname{Var}\left(\hat{\gamma}_{1}+\hat{\gamma}_{2}+\hat{\gamma}_{3}+\hat{\alpha}_{1} \hat{\theta}_{V}\right) \\
= & \operatorname{Var}\left(\hat{\gamma}_{1}\right)+\operatorname{Var}\left(\hat{\gamma}_{2}\right)+\operatorname{Var}\left(\hat{\gamma}_{3}\right) \\
& +\operatorname{Var}\left(\hat{\alpha}_{1}\right) \operatorname{Var}\left(\hat{\theta}_{V}\right)+2 \alpha_{1} \operatorname{Var}\left(\hat{\theta}_{V}\right)+2 \theta_{V} \operatorname{Var}\left(\hat{\alpha}_{1}\right) \\
& +2 \operatorname{Cov}\left(\hat{\gamma}_{1}, \hat{\gamma}_{2}\right)+2 \operatorname{Cov}\left(\hat{\gamma}_{1}, \hat{\gamma}_{3}\right)+2 \operatorname{Cov}\left(\hat{\gamma}_{2}, \hat{\gamma}_{3}\right) \\
& +2 \alpha_{1} \operatorname{Cov}\left(\hat{\gamma}_{1}, \hat{\theta}_{V}\right)+2 \theta_{V} \operatorname{Cov}\left(\hat{\gamma}_{1}, \hat{\alpha}_{1}\right)+2 \alpha_{1} \operatorname{Cov}\left(\hat{\gamma}_{2}, \hat{\theta}_{V}\right) \\
& +2 \theta_{V} \operatorname{Cov}\left(\hat{\gamma}_{2}, \hat{\alpha}_{1}\right)+2 \alpha_{1} \operatorname{Cov}\left(\hat{\gamma}_{3}, \hat{\theta}_{V}\right)+2 \theta_{V} \operatorname{Cov}\left(\hat{\gamma}_{3}, \hat{\alpha}_{1}\right).
\end{aligned}
$$
Assuming asymptotic normality, the $95 \%$ confidence interval of ATE would be computed as
$$
\left[\hat{\delta}_{A T E}-1.96 \sqrt{\operatorname{Var}\left(\hat{\delta}_{A T E}\right)}, \hat{\delta}_{A T E}+1.96 \sqrt{\operatorname{Var}\left(\hat{\delta}_{A T E}\right)}\right].
$$
However, without empirical information of $\operatorname{Cov}\left(\hat{\gamma}_{1}, \hat{\alpha}_{1}\right), \operatorname{Cov}\left(\hat{\gamma}_{2}, \hat{\alpha}_{1}\right)$, and $\operatorname{Cov}\left(\hat{\gamma}_{3}, \hat{\alpha}_{1}\right)$, the analyst may ignore the estimation uncertainty in the Stage 1 analysis by viewing $\widehat{\alpha}_{1}$ as a constant. Additionally, because the sample estimates $\hat{\alpha}_{1 k}, \hat{\beta}_{1 k}$, and $\hat{\beta}_{2 k}$ obtained from the Stage 1 analysis are used as predictors in the Stage 2 analysis, estimators of the coefficients in the Stage 2 regression $\hat{\gamma}_{1}, \hat{\gamma}_{2}, \hat{\gamma}_{3}$, and $\hat{\theta}_{V}$ and their variances and covariances likely contain bias. These issues potentially jeopardize the statistical inference for $\delta_{\text {ATE }}$ \citep{CameronTrivedi2005}.

\textit{Multilevel bootstrap BCa.} To account for Stage 1 estimation uncertainty in two-stage analyses, researchers often resort to bootstrapping for robust inference. In this study, we adopt a multilevel bootstrapping approach \citep{Goldstein2011}. We first generate a bootstrap sample of sites through a simple random resampling with replacement of the site indices. We then generate a bootstrap sample of individuals within each treatment group at each bootstrapped site. By repeating this procedure, we generate a total of $B$ bootstrapped samples and obtain $B$ bootstrapped estimates of $\delta_{A T E}$. Rather than using the conventional percentile method to construct the $95 \%$ confidence interval, we opt for the bias-corrected accelerated (BCa) method. The latter adjusts for both bias and skewness in the bootstrap distribution and allows the standard error to be a function of the parameter value. Therefore, BCa is second-order accurate and is preferrable to the percentile method that is only first-order accurate \citep{EfronTibshirani1993}.

\section{Simulations}\label{sec3}
\subsection{Research Questions for the Simulations}
We empirically address the following two questions through simulations. The first question is how the proposed MS2T-IV strategy compares to existing alternative methods in estimating the cumulative ATE; the second question is how the coverage rates compare between the improper inference strategy and the multiple bootstrap BCa strategy. In addition, we examine through simulations the consequences when Assumption 2 and 3 are violated.

We have proposed the MS2T-IV method, both with and without covariance adjustment for site mean-centered baseline covariates $X$. Alternative methods include the IPTW method and a naive approach. IPTW adjusts for observed baseline covariates $X$ and time-varying covariate $V$ through weighting. The naive method regresses $Y$ on $Z$, $D$, and $Z$-by-$D$ interaction while accounting for site membership, with or without covariance adjustment for $X$. We do not include the standard IV method or the MSMM-IV method for comparisons, as they are not designed to estimate the cumulative ATE. Evaluation criteria include bias, efficiency, mean-squared-error, and degree of deviation from the nominal coverage rate.

When Assumptions \ref{ass1}-\ref{ass3} as well as the model-based assumptions hold, the MS2T-IV estimator is expected to remove bias not only associated with $X$ and $V$ but also with unobserved baseline covariates $U$. Covariate adjustment for the site mean-centered $X$ is expected to further reduce finite-sample bias and improve efficiency. We hypothesize that, as the number of sites and the sample size per site increase, both the bias and the empirical variance of the MS2T-IV estimator are expected to converge to zero.

\subsection{Data Generation}
We generate simulation data in accordance with the theoretical model (\ref{eq1}). Baseline covariates include an observed $X$ and an unobserved $U$. Following typical multisite designs in application research, we specify the number of sites to be either 25 or 100 and let the sample size per site vary from 30 to 5,000. We generate 500 independent data sets in each case. Details of the data generation are in section S3 of the supplementary material.

\subsection{Simulation Results: Estimation}
The simulation results for estimation performance are summarized in Table \ref{table2}.

\textit{Bias.} By employing the instrumental variable to remove confounding associated with $U$ in addition to that associated with $X$ and $V$, the MS2T-IV estimator contains relatively less bias in comparison with the alternative strategies. Consistent with our hypothesis, covariance adjustment for site mean-centered $X$ in the MS2T-IV analysis further reduces bias. As the sample size per site increases, the MS2T-IV estimator shows a clear downward trend in bias. In contrast, the IPTW estimator fails to remove the bias associated with $U$ while the naive estimator with covariance adjustment for $X$ fails to remove the bias associated with $U$ and $V$. Such bias remains despite an increase in the sample size per site. Across all these methods, increasing the number of sites does not necessarily reduce bias.

\textit{Efficiency.} Across all the methods under consideration, the empirical variance decreases when either the number of sites or the sample size per site increases. The MS2T-IV estimator is relatively less efficient than the naive estimator but relatively more efficient than the IPTW estimator. For the naive estimator and the MS2T-IV estimator, covariance adjustment for $X$ improves efficiency to a limited extent.

\textit{MSE.} Among all the estimators compared, only the MS2T-IV estimators are consistent and show a fast rate of MSE reduction when the number of sites or the sample size per site increases. In contrast, because the naive estimators and the IPTW estimator are biased, their MSE does not converge to zero.

When the within-site zero covariance assumption (Assumption 2) is violated, none of the five estimators will be consistent. The bias does not converge to zero as the sample size per site increases. When the between-site independence assumption (Assumption 3) is violated, the clear downward trend in bias also disappears for the MS2T-IV estimator. Nevertheless, in these cases, the MS2T-IV estimator with adjustment for $X$ still almost always has the smallest bias and MSE among all five estimators considered in this simulation study. Details of the simulation results with violated assumptions are in section S5 of the supplementary material.

\begin{table}[ht]
\centering
\caption{Estimation Performance}
\label{table2}
\begin{tabular}{@{}lcrrrrrrrr@{}}
\toprule
&& \multicolumn{4}{c}{$K=25$} & \multicolumn{4}{c}{$K=100$} \\
\cmidrule(lr){3-6}\cmidrule(lr){7-10}
& $n_k$ & 30 & 100 & 1{,}000 & 5{,}000 & 30 & 100 & 1{,}000 & 5{,}000 \\
\midrule
\multicolumn{9}{@{}l}{\textbf{Bias}} \\
Naive & & 1.79 & 1.75 & 1.80 & 1.80 & 1.73 & 1.74 & 1.79 & 1.75 \\
Naive-adj & & 1.35 & 1.32 & 1.38 & 1.38 & 1.23 & 1.33 & 1.38 & 1.33 \\
IPTW & & -2.52 & -1.45 & -0.50 & -0.37 & -2.69 & -1.45 & -0.58 & -0.49 \\
MS2T-IV & & -0.05 & -0.03 & 0.03 & -0.01 & -0.06 & -0.03 & 0.03 & 0.01 \\
MS2T-IV-adj & & -0.09 & -0.07 & 0.02 & -0.01 & -0.10 & -0.07 & 0.01 & 0.01 \\
\addlinespace[0.6ex]
\multicolumn{9}{@{}l}{\textbf{Empirical Variance}} \\
Naive & & 5.34 & 2.72 & 0.97 & 0.91 & 1.44 & 0.58 & 0.25 & 0.24 \\
Naive-adj & & 4.86 & 2.35 & 0.77 & 0.75 & 1.42 & 0.51 & 0.21 & 0.19 \\
IPTW & & 25.33 & 9.04 & 4.93 & 4.32 & 6.73 & 2.24 & 1.27 & 1.24 \\
MS2T-IV & & 8.90 & 4.29 & 1.69 & 1.64 & 2.08 & 1.15 & 0.44 & 0.31 \\
MS2T-IV-adj & & 8.13 & 3.58 & 1.62 & 1.61 & 1.94 & 1.01 & 0.41 & 0.31 \\
\addlinespace[0.6ex]
\multicolumn{9}{@{}l}{\textbf{MSE}} \\
Naive & & 8.53 & 5.78 & 4.20 & 4.15 & 4.43 & 3.60 & 3.46 & 3.29 \\
Naive-adj & & 6.68 & 4.09 & 2.66 & 2.65 & 2.94 & 2.28 & 2.10 & 1.96 \\
IPTW & & 31.67 & 11.13 & 5.18 & 4.45 & 13.96 & 4.36 & 1.60 & 1.48 \\
MS2T-IV & & 8.91 & 4.30 & 1.70 & 1.64 & 2.08 & 1.15 & 0.44 & 0.31 \\
MS2T-IV-adj & & 8.13 & 3.58 & 1.62 & 1.61 & 1.95 & 1.02 & 0.41 & 0.31 \\
\bottomrule
\end{tabular}

\vspace{0.25ex}
\raggedright\footnotesize
Note: $K$ is the number of sites. $n_k$ is the sample size per site.
\end{table}

\subsection{Simulation Results: Statistical Inference}
To address the research question on statistical inference, we focus on the MS2T-IV estimator with covariance adjustment. From the 500 simulations in each scenario, we observe that the distribution of this estimator is approximately normal. Table \ref{table3} compares the coverage rates between the two different inference methods. We have included a scenario that resembles the number of sites $(K=76)$ and the sample size per site $\left(n_{k}=60\right)$ in the current real data application.

The improper method for statistical inference shows under-coverage across all scenarios, suggesting that ignoring the Stage 1 estimation uncertainty typically results in an underestimation of the standard error for the MS2T-IV estimator. In contrast, the bootstrap BCa method shows over-coverage in most scenarios including the one that resembles the current real data application. Therefore, we should note that statistical inference based on the improper method is likely too liberal, while inference based on the Bootstrap BCa method tends to be overly conservative.

\begin{table}[ht]
\centering
\caption{Coverage Rates}
\label{table3}
\begin{tabular}{cccc}
\toprule
$K$ & $n_k$ & {\text{Improper Method}(\%)} & {\text{Bootstrap BCa}~(\%)} \\
\midrule
76 & 60 & 94.0 & 98.4 \\
\midrule
25 & 30 & 91.6 & 99.6 \\
25 & 100 & 93.8 & 99.2 \\
25 & 1{,}000 & 93.6 & 97.4 \\
100 & 30 & 93.2 & 99.2 \\
100 & 100 & 93.8 & 98.4 \\
100 & 1{,}000 & 93.2 & 95.6 \\
\bottomrule
\end{tabular}
\end{table}

\section{Application Study}\label{sec4}
We illustrate the MS2T-IV method with an evaluation of the impact of class size reduction in the early grades on student learning. The kindergarten sample in Project STAR consisted of 6,325 students across 79 participating schools (the Project STAR public dataset is available online at \href{https://search.rproject.org/CRAN/refmans/AER/html/STAR.html}{https://search.rproject.org/CRAN/refmans/AER/html/STAR.html}). At the beginning of the kindergarten year, students were randomly assigned within each school to either a small or regular class; and kindergarten teachers were randomly assigned to these classes as well. Due to family relocations, 1,810 kindergartners left Project STAR after one year, while 2,313 new students enrolled in the participating schools. Among the students who left after kindergarten, those who attended a small class and those in a regular class did not differ in observed pretreatment characteristics. Students who were newly enrolled in Grade 1 were randomly assigned to either a small or regular class, with Grade 1 teachers also being randomly assigned to these classes.

Our causal estimand is the cumulative ATE of attending a small class versus a regular class over the two consecutive years. The target population for this study is represented by the 4,511 students who were enrolled in 76 Project STAR schools in both kindergarten and Grade 1. However, noncompliance started to occur at the beginning of Grade 1: 108 students switched from a small class to a regular class, while 248 students switched from a regular class to a small class at the end of the kindergarten year. The noncompliance rate was about $8 \%$ in each treatment group. \citet{Krueger1999} speculated that these moves between class size types were likely due to behavioral problems or parental complaints.

The outcome measure is the sum of reading and math test scores by the end of Grade 1, while the sum of reading and math test scores by the end of kindergarten is a key posttreatment confounder in this study. Table \ref{table4} compares the actual class sizes and achievement scores in kindergarten and Grade 1 across the four treatment sequences. A naive comparison between students who stayed in a small class and those who stayed in a regular class over both kindergarten and Grade 1 shows a 24-point difference in Grade 1 achievement. Yet based on an analysis regressing the kindergarten score on the $Z$-by-$D$ group dummies with school fixed effects, students who switched from a small class to a regular class ($Z=1$, $D=0$) displayed significantly lower kindergarten scores on average compared to those who remained in a small class $(Z=1, D=1)$. Similarly, students who switched from a regular class to a small class ($Z=0, D=1$) displayed lower kindergarten scores than those who remained in a regular class ($Z=0, D=0$). The average between-group difference is $-23.5$ with a p-value less than 0.001 in the former case, and $-1.6$ with a p-value of 0.345 in the latter case. Hence the naive comparison between the ($Z= 1, D=1)$ group and the ($Z=0, D=0$) group would likely contain a positive bias. Further analysis reveals that the sum of kindergarten reading and math scores is a statistically significant predictor of compliance among those in the $Z=1$ group only.

\begin{table}[ht]
\centering
\caption{Kindergarten and Grade 1 Actual Class Sizes and Test Scores by Treatment Sequences}
\label{table4}
\begin{tabular}{@{}l cc cc cc cc @{}}
\toprule
& \multicolumn{2}{c}{$Z=0, D=0$} & \multicolumn{2}{c}{$Z=0, D=1$} & \multicolumn{2}{c}{$Z=1, D=0$} & \multicolumn{2}{c}{$Z=1, D=1$} \\
& \multicolumn{2}{c}{$(N=2{,}867)$} & \multicolumn{2}{c}{$(N=248)$} & \multicolumn{2}{c}{$(N=108)$} & \multicolumn{2}{c}{$(N=1{,}292)$} \\
\cmidrule(lr){2-3}\cmidrule(lr){4-5}\cmidrule(lr){6-7}\cmidrule(lr){8-9}
& Mean & SD & Mean & SD & Mean & SD & Mean & SD \\
\midrule
K Class Size & 22 & 2 & 22 & 2 & 15 & 1 & 15 & 2 \\
G1 Class Size & 23 & 2 & 16 & 2 & 22 & 3 & 16 & 2 \\
K Reading + Math & 928 & 70 & 921 & 75 & 919 & 55 & 943 & 74 \\
G1 Reading + Math & 1{,}055 & 91 & 1{,}060 & 89 & 1{,}049 & 92 & 1{,}079 & 95 \\
\bottomrule
\end{tabular}
\end{table}

The pretreatment covariates ($X$) include student gender, race, age at kindergarten entry, free-lunch status, repetition status in kindergarten, special education status in kindergarten, and urbanicity. For any missing observations in the discrete covariates, we create a separate missing category. For missing values in the continuous covariates, we use the ``mice" package in R to conduct multiple imputation, generating ten imputed datasets.

We then apply the MS2T-IV strategy to each imputed dataset. At Stage 1, we obtain OLS estimates of the school-specific ITT effects. To increase the estimation efficiency and reduce bias, we adjust for the pretreatment covariates $X$ each centered at its school mean. The estimated school-specific ITT effects of $Z$ on $V, D$, and $Y$ in school $k$ are denoted respectively as $\hat{\alpha}_{1 k}, \hat{\beta}_{1 k}$, and $\hat{\theta}_{1 k}$, for $k=1, \ldots, K$. We also estimate the fraction of students who remain in a small class among those initially assigned to a small class in school $k$, which is denoted as $\hat{\beta}_{2 k}$. At Stage 2, we conduct a school-level analysis, regressing $\hat{\theta}_{1 k}$ on $\hat{\alpha}_{1 k}, \hat{\beta}_{1 k}$, and $\hat{\beta}_{2 k}$ as specified earlier in model (\ref{eq6}). Following equation (\ref{eq7}), we compute the point estimate and further obtain interval estimates of the cumulative ATE.

According to our analytic results aggregated over the ten imputed datasets, attending a small class as opposed to a regular class in both kindergarten and Grade 1 is expected to increase the sum of the Grade 1 reading and math scores by 20.21 points. Using the within-school standard deviation of the outcome in the control group as the scaling unit, the effect size is about 0.26. The improper $95 \%$ confidence interval for the cumulative ATE is $(6.91,33.51)$ while the Bootstrap BCa confidence interval is $(5.67,77.39)$. In the latter case, we generated 500 bootstrapped samples. For each bootstrapped sample, we calculated the average estimate of the cumulative ATE across the 10 imputed datasets. Based on the 500 bootstrapped estimates, we constructed the Bootstrap BCa confidence interval.

\section{Discussion}\label{sec5}
\subsection{Contributions of this study}
Interventions that span two or more phases may have the potential to generate long-lasting impacts. However, noncompliance may arise after participants have already received substantial exposure to the treatment in the first phase. This study addresses a key challenge in evaluating such programs: identifying the cumulative average treatment effect of a two-phase intervention when noncompliance occurs after Phase I. Such noncompliance is often linked to an individual's intermediate response to the Phase I treatment, along with other selection factors. This methodological advancement is essential for determining whether an additional phase of treatment may enhance overall effectiveness.

Capitalizing on multisite randomized trials, the proposed MS2T-IV strategy consistently estimates the cumulative ATE of a two-phase treatment sequence versus a control sequence. This strategy relies on a set of identification assumptions that are arguably weaker than those required by existing methods.

First, by using the Phase I random treatment assignment within each site as an instrumental variable, this strategy relaxes the conventional sequential ignorability assumption by allowing for unobserved pretreatment confounders in the relationship between the Phase II treatment and the outcome.

Second, this strategy explicitly models the observed intermediate outcomes of the Phase I treatment as posttreatment confounders that predict both the Phase II treatment and the outcome, and it additionally allows for a direct effect of the Phase I treatment on the outcome, thereby relaxing the exclusion restriction.

We can show that in a multisite IV analysis, omitting a posttreatment confounder $V$ may introduce bias in identifying the cumulative ATE. This bias will be zero only if the ITT effect of $Z$ on $V$ does not covary with the ITT effect of $Z$ on $D$. This condition holds if either $V$ does not predict $D$ or if the ITT effect of $Z$ on $V$ is constant across sites. Section S2 of the supplementary material derives the bias when $V$ is an omitted posttreatment confounder. This result enables analysts to conduct a sensitivity analysis to assess the potential bias associated with such an omission.

According to the simulation results, in comparison with the naive method that adjusts only for observed pretreatment covariates and with the IPTW method that adjusts for observed pretreatment and posttreatment covariates, the MS2T-IV method excels in bias
reduction. Although the MS2T-IV estimator is generally less efficient than these existing alternatives, it outperforms them in MSE reduction when the number of sites is sufficiently large.
\vspace{-0.5em}
\subsection{Limitations and Future Research}
Several limitations of the MS2T-IV method are to be overcome through future research.

First, the MS2T-IV method and other existing methods for identifying the cumulative ATE share a common limitation: none are equipped to handle unobserved posttreatment confounders. If an unobserved posttreatment confounder $L$ exists even after conditioning on $V$, the MS2T-IV estimator is likely biased due to the omission of $L$, unless either $L$ does not predict $D$ conditioning on $V$, or the ITT effect of $Z$ on $L$ is constant across sites.

Second, as suggested by our simulation results, adjusting for site mean-centered pretreatment covariates may reduce finite-sample bias and improve estimation efficiency. Future research may further quantify the remaining finite-sample bias and explore strategies for bias correction.

Third, our assumptions 2 and 3 are violated if there are individual-level or site-level modifiers. So if the coefficients differ by subpopulations of individuals or subpopulations of sites, then these assumptions are likely violated. In such cases, the analysis will need to be conducted within each subpopulation. And eventually, the cumulative ATE will need to be an average over the subpopulations.
Yet even if all the modifiers are observed, this may be impractical due to sample size constraints.

Fourth, it is not uncommon that a treatment sequence may consist of more than two phases. Noncompliance may occur in multiple phases including phase I. Future research may extend the proposed strategy to evaluations of the cumulative ATE of multi-phase interventions.

\section{Competing interests}
No competing interest is declared.

\vspace{-0.5cm}
\section{Author contributions statement}
Guanglei Hong, Xu Qin, Zhengyan Xu, and Fan Yang have made equal contributions and are listed alphabetically.

\vspace{-0.5cm}
\section{Acknowledgments}
The research reported here was supported by the Institute of Education Sciences, U.S. Department of Education, through a Statistical and Research Methodology Grant (R305D120020). This material is also based upon work supported by the National Science Foundation under Grant Number 2337612 in the form of a CAREER Award for the second author. Any opinions, findings, and conclusions or recommendations expressed in this material are those of the authors and do not necessarily reflect the views of the Institute of Education Sciences or the National Science Foundation. The authors report no competing interests to declare.

\bibliographystyle{abbrvnat}
\bibliography{reference}

\newpage

\begin{center}
{\bf Supplementary Material for 
``Two-Phase Treatment with Noncompliance: Identifying the Cumulative Average Treatment Effect via Multisite Instrumental Variables"}
\end{center}
\if0\blind
{}\fi

\setcounter{page}{1}
\setcounter{equation}{0}
\setcounter{section}{0}
\setcounter{figure}{0}
\setcounter{table}{0}
\renewcommand {\thepage} {S\arabic{page}}
\renewcommand {\theequation} {S\arabic{equation}}
\renewcommand {\thesection} {S\arabic{section}}
\renewcommand{\thefigure}{S\arabic{figure}}
\renewcommand{\thetable}{S\arabic{table}}

Section S1 presents the derivations of the reduced-form models.

Section S2 presents the derivation of the bias resulting from the omission of a posttreatment confounder $V$.

Section S3 outlines the data generation plan for the simulations.

Section S4 describes the alternative methods for comparison.

Section S5 displays the simulation results when the assumptions are violated.

\section{Derivations of the Reduced-Form Models}


\subsection{Reduced-Form Model for $\boldsymbol{Y}$}
Here we derive the results for the reduced-form model for $Y$. For each school $k$, following the theoretical model for $Y_{i k}(z, d)$, we have that

\begin{align*}
E\left\{Y_{i k}(0) \mid S_{i}=k\right\}&=E\left\{Y_{i k}\left(0, D_{i k}(0)\right) \mid S_{i}=k\right\}\\&=E\left\{\theta_{Y 0 i k}+\theta_{V i k} V_{i k}(0)+\gamma_{2 i k} D_{i k}(0) \mid S_{i}=k\right\}, 
\end{align*}

and 

\begin{align*}
E\left\{Y_{i k}(1) \mid S_{i}=k\right\}&=E\left[Y_{i k}\left(1, D_{i k}(1)\right) \mid S_{i}=k\right] \\
&=E\left\{\theta_{Y 0 i k}+\theta_{V i k} V_{i k}(1)+\gamma_{1 i k}+\gamma_{2 i k} D_{i k}(1)+\gamma_{3 i k} D_{i k}(1) \mid S_{i}=k\right\}.
\end{align*}

Then, the ITT effect of $Z$ on $Y$ at site k is
\begin{align*}
&~~~~E\left\{Y_{i k}(1)-Y_{i k}(0) \mid S_{i}=k\right\}\\&=E\left[\theta_{V i k} \left\{V_{i k}(1)-V_{i k}(0)\right\}+\gamma_{1 i k}+\gamma_{2 i k} \left\{D_{i k}(1)-D_{i k}(0)\right\}+\gamma_{3 i k} D_{i k}(1) \mid S_{i}=k\right]\\&=\theta_{V k} \alpha_{1 k}+\gamma_{1 k}+\gamma_{2 k} \beta_{1 k}+\gamma_{3 k} \beta_{2 k},
\end{align*}
under the assumption 2 that $\mathrm{Cov}\{\gamma_{2 i k}, \beta_{1ik}\} = \mathrm{Cov}\{\gamma_{3 i k}, \beta_{2ik}\}=\mathrm{Cov}\{\theta_{Vik}, \alpha_{1ik}\}$.

Given that $Z_{ik}$ is randomized, the analytic model for obtaining the ITT effect on $Y$ at site $k$ is
$$
\begin{aligned}
&Y_{i k}=\theta_{0 k}+\theta_{1 k} Z_{i k}+e_{Y i k},
\end{aligned}
$$
where $\theta_{0 k}=E\left\{Y_{i k}(0) \mid S_{i}=k\right\};$
$\theta_{1 k}=\theta_{V k} \alpha_{1 k}+\gamma_{1 k}+\gamma_{2 k} \beta_{1 k}+\gamma_{3 k}\beta_{2k} ;$ and $e_{Y i k}$ has mean zero and is uncorrelated with $Z_{i k}$ within school $k$.

\subsection{School-level population model}
Here, we show that $\epsilon_k$ defined in model (6) has mean zero and is uncorrelated with the predictors $\beta_{1k}, \beta_{2k}$ and $\alpha_{1k}$.

Since $\epsilon_{k}=\left(\gamma_{1 k}-\gamma_{1}\right)+\left(\gamma_{2 k}-\gamma_{2}\right) \beta_{1 k}+\left(\gamma_{3 k}-\gamma_{3}\right) \beta_{2 k}+\left(\theta_{V k}-\theta_{V}\right) \alpha_{1 k}$, and therefore,
\begin{align*}\mathrm{E}(\epsilon_{k})&=\mathrm{E}\left(\gamma_{1 k}-\gamma_{1}\right)+\mathrm{E}\{\left(\gamma_{2 k}-\gamma_{2}\right) \beta_{1 k}\}+\mathrm{E}\{\left(\gamma_{3 k}-\gamma_{3}\right) \beta_{2 k}\}+\mathrm{E}\{\left(\theta_{V k}-\theta_{V}\right) \alpha_{1 k}\}\\&=0+\mathrm{E}\left(\gamma_{2 k}-\gamma_{2}\right)\mathrm{E}(\beta_{1 k})+\mathrm{E}\left(\gamma_{3 k}-\gamma_{3}\right) \mathrm{E}(\beta_{2 k})+\mathrm{E}\left(\theta_{V k}-\theta_{V}\right) \mathrm{E}(\alpha_{1 k})\\&=0,
\end{align*}
where the second identity holds under Assumption 3.

\begin{align*}
\mathrm{Cov}(\epsilon_k, \beta_{1k})
&=\mathrm{Cov}(\gamma_{1k}-\gamma_1, \beta_{1k})+\mathrm{Cov}\{(\gamma_{2k}-\gamma_2)\beta_{1k}, \beta_{1k}\}+\mathrm{Cov}\{(\gamma_{3k}-\gamma_3)\beta_{2k},\beta_{1k}\}
\\&+\mathrm{Cov}\{(\theta_{Vk}-\theta_{V})\alpha_{1k},\beta_{1k}\}.\end{align*}
In the above expression, $\mathrm{Cov}(\gamma_{1k}-\gamma_1, \beta_{1k})=0$ given the independence between $\gamma_{1k}$ and $\beta_{1k}$ under Assumption 3.
The term $\mathrm{Cov}\{(\gamma_{2k}-\gamma_2)\beta_{1k}, \beta_{1k}\} =0$ given the independence between $\gamma_{2k}$ and $\beta_{1k}$ under Assumption 3.
The term $\mathrm{Cov}\{(\gamma_{3k}-\gamma_3)\beta_{2k},\beta_{1k}\} = 0$ given the joint independence between $(\beta_{1k}, \beta_{2k})$ with $\gamma_{3k}$ under Assumption 3. The term $\mathrm{Cov}\{(\theta_{Vk}-\theta_{V})\alpha_{1k},\beta_{1k}\}=0$ given the joint independence between $(\alpha_{1k}, \beta_{1k})$ with $\theta_{Vk}$ under Assumption 3.

Similarly, one can show that $\mathrm{Cov}(\epsilon_k, \beta_{2k})$ and $\mathrm{Cov}(\epsilon_k, \alpha_{1k})$ are both zero under Assumption 3.

\section{Derivation of Bias due to the Omission of Postreatment Confounder V}\label{appB}
If we omit the posttreatment confounder $V$, the Stage 1 analysis would provide $\beta_{1 k}, \beta_{2 k}$ and $\theta_{1 k}$, but not $\alpha_{1 k}$. In addition, the Stage 2 model would become:
$$
\theta_{1 k}=\gamma_{1}^{\prime}+\gamma_{2}^{\prime} \beta_{1 k}+\gamma_{3}^{\prime} \beta_{2 k}+\epsilon_{k}^{\prime} .
$$

A derivation based on the reduced-form models suggests that the cumulative ATE is equal to $\gamma_{1}^{\prime}+\gamma_{2}^{\prime}+\gamma_{3}^{\prime}$. Now we derive the bias in $\gamma_{1}^{\prime}+\gamma_{2}^{\prime}+\gamma_{3}^{\prime}$ when the true $\delta_{A T E}=\gamma_{1}+\gamma_{2}+\gamma_{3}+ \theta_{V} \alpha_{1}$.

Without controlling for $\alpha_{1 k}$ in the Stage 2 analysis, the bias in $\gamma_{2}^{\prime}$ and that in $\gamma_{3}^{\prime}$ are given by:
$$
\begin{aligned}
\gamma_{2}^{\prime}-\gamma_{2} & =\frac{\theta_{V} \operatorname{Cov}\left(\beta_{1 k}, \alpha_{1 k}\right)\left\{\operatorname{Var}\left(\beta_{2 k}\right)+\operatorname{Cov}\left(\beta_{1 k}, \beta_{2 k}\right)\right\}}{\operatorname{Var}\left(\beta_{1 k}\right) \operatorname{Var}\left(\beta_{2 k}\right)-\left\{\operatorname{Cov}\left(\beta_{1 k}, \beta_{2 k}\right)\right\}^{2}}, \\
\gamma_{3}^{\prime}-\gamma_{3} & =\frac{\theta_{V} \operatorname{Cov}\left(\beta_{2 k}, \alpha_{1 k}\right)\left\{\operatorname{Var}\left(\beta_{1 k}\right)+\operatorname{Cov}\left(\beta_{1 k}, \beta_{2 k}\right)\right\}}{\operatorname{Var}\left(\beta_{1 k}\right) \operatorname{Var}\left(\beta_{2 k}\right)-\left\{\operatorname{Cov}\left(\beta_{1 k}, \beta_{2 k}\right)\right\}^{2}} .
\end{aligned}
$$
Given that $E\left(\theta_{1 k}\right)=\gamma_{1}^{\prime}+\gamma_{2}^{\prime} \beta_{1}+\gamma_{3}^{\prime} \beta_{2}=\gamma_{1}+\gamma_{2} \beta_{1}+\gamma_{3} \beta_{2}+\theta_{V} \alpha_{1}$ where $\beta_{1}=E\left(\beta_{1 \mathrm{k}}\right)$ and $\beta_{2}=E\left(\beta_{2 \mathrm{k}}\right)$, we have that
$$
\gamma_{1}^{\prime}-\gamma_{1}=\left(\gamma_{2}-\gamma_{2}^{\prime}\right) \beta_{1}+\left(\gamma_{3}-\gamma_{3}^{\prime}\right) \beta_{2}+\theta_{V} \alpha_{1} .
$$
Based on the above calculation, we have that
$$
\left(\gamma_{1}^{\prime}+\gamma_{2}^{\prime}+\gamma_{3}^{\prime}\right)-\left(\gamma_{1}+\gamma_{2}+\gamma_{3}+\theta_{V} \alpha_{1}\right)=\left(\gamma_{2}-\gamma_{2}^{\prime}\right)\left(1-\beta_{1}\right)+\left(\gamma_{3}-\gamma_{3}^{\prime}\right)\left(1-\beta_{2}\right) .
$$
This is the bias when the posttreatment confounder $V$ is omitted. The bias is 0 if $\gamma_{2}= \gamma_{2}^{\prime}$ and $\gamma_{3}=\gamma_{3}^{\prime}$, which occurs when $\theta_{V}=0$ or when $\operatorname{Cov}\left(\beta_{1 k}, \alpha_{1 k}\right)=\operatorname{Cov}\left(\beta_{2 k}, \alpha_{1 k}\right)=0$. However, as a posttreatment covariate, $V$'s impact on $Y$ (i.e., $\theta_{V}$ ) is expected to be nonzero. In addition, because $V$ lies on the pathway from $Z$ to $D, \alpha_{1 k}$ contributes to both $\beta_{1 k}$ and $\beta_{2 k}$, and is therefore correlated with them unless $\alpha_{1 k}$ is constant across all sites. Thus, we can conclude that omitting a posttreatment confounder would introduce bias.

The above derivation has revealed that, to conduct a sensitivity analysis when a posttreatment confounder is omitted, the key sensitivity parameters are $\theta_{V}, \operatorname{Cov}\left(\beta_{1 k}, \alpha_{1 k}\right)$, and $\operatorname{Cov}\left(\beta_{2 k}, \alpha_{1 k}\right)$. Increases in the magnitude of any of these sensitivity parameters will amplify the differences $\gamma_{2}-\gamma_{2}^{\prime}$ or $\gamma_{3}-\gamma_{3}^{\prime}$, thereby potentially increasing the magnitude of the bias.

\section{Data Generation Plan for the Simulations}

For the simulation study, we generate 500 independent data sets. In each data set, there are $K$ sites and $n_k$ individuals per site. Let $K = 25, 100$, and $n_k = 30, 100, 1,000, 5,000$.

Step 1. We generate the Phase I treatment assignment $Z_{ik}$. Let $P_k^{(Z)}$ denote the probability of $Z_{ik} = 1$ in site $k$ for $k = 1, \dots, K$. Let $P_k^{(Z)} \sim \text{Uniform}(0.25, 0.35)$. Within site $k$, $P_k^{(Z)}$ is constant for all individuals. We randomly assign $P_k^{(Z)}n_k$ individuals to the Phase I treated group ($Z_{ik}=1$) in site $k$. This process will resemble the experimental design in the real-world application.

Step 2. We then generate a binary $U_{ik}$. Let $P_k^{(U)}$ denote the probability that $U_{ik}=1$ in site $k$ for $k=1, \dots, K$. Let $P_k^{(U)}\sim \text{Uniform}(0.25,0.45)$. Let $P_{ik}^{(U)}$ denote the probability that $U_{ik}=1$ for individual $i$ in site $k$, which is uniformly distributed around the site mean: $P_{ik}^{(U)}\sim \text{Uniform}(P_k^{(U)}-0.02, P_k^{(U)}+0.02)$. Finally, we generate $U_{ik}$ by taking random draws from Bernoulli distributions with parameter value $P_{ik}^{(U)}$.

Step 3. We also generate a binary $X_{ik}$. Let $P_k^{(X)}$ denote the probability that $X_{ik}=1$ in site $k$ for $k=1, \dots, K$. Let $P_k^{(X)}\sim \text{Uniform}(0.3,0.5)$. Let $P_{ik}^{(X)}$ denote the probability that $X_{ik}=1$ for individual $i$ in site $k$, which is uniformly distributed around the site mean: $P_{ik}^{(X)}\sim \text{Uniform}(P_k^{(X)}-0.02, P_k^{(X)}+0.02)$. Finally, we generate $X_{ik}$ by taking random draws from Bernoulli distributions with parameter value $P_{ik}^{(X)}$.

Step 4. We generate the Phase I outcome $V_{ik}(z)$ for $z=0,1$, each as a function of school-mean-centered $U_{ik}$ and $X_{ik}$ with different site-specific intercepts. To simplify, the slopes for $U_{ik}$ and $X_{ik}$ are constant across the sites, and to satisfy the within-site zero covariance assumption (Assumption 2), they do not differ for $V_{ik}(0)$ and $V_{ik}(1)$.
$$
\begin{aligned}
    V_{ik}(0) &= 35 + t_{0k} + 10(X_{ik}-\bar X_k) + 20(U_{ik}-\bar U_k), \\
    V_{ik}(1) &= 40 + t_{0k} + t_{1k} + 10(X_{ik}-\bar X_k) + 20(U_{ik}-\bar U_k),
\end{aligned}
$$
where $t_{0k} \sim N(0,8^2)$ and $t_{1k} \sim N(0,6^2)$ are independent.

The corresponding parameters in our theoretical model are as follows:
$$
V_{ik}(z) = \alpha_{0ik} + \alpha_{1ik} z,
$$
$$
\begin{aligned}
    \alpha_{0ik} &= 35 + t_{0k} + 10(X_{ik}-\bar X_k) + 20(U_{ik}-\bar U_k), \\
    \alpha_{1ik} &= 5 + t_{1k}.
\end{aligned}
$$

The observed Phase I outcome is $V_{ik} = Z_{ik}V_{ik}(1) + (1 - Z_{ik})V_{ik}(0)$.

Step 5. To generate the Phase II treatment assignment $D_{ik}(z)$ for $z=0,1$, we specify a pair of logit models as follows:
$$
\begin{aligned}
    D_{ik}(0) &= \mathbbm{1}\left\{-(X_{ik}-\bar X_k) - (U_{ik}-\bar U_k) - 0.1V_{ik}(0) + s_{0k} - u_{ik}^{(0)} \ge 0\right\}, \\
    D_{ik}(1) &= \mathbbm{1}\left\{X_{ik}-\bar X_k + U_{ik}-\bar U_k + 0.05V_{ik}(1) + s_{1k} - u_{ik}^{(1)} \ge 0\right\},
\end{aligned}
$$
where $s_{0k}\sim N(0,1)$ and $s_{1k}\sim N(0,1)$ are independent and each i.i.d. across sites, $u_{ik}^{(0)}$ and $u_{ik}^{(1)}$ are independent i.i.d. Logistic(0, 1) random variables.

The observed Phase II treatment is $D_{ik} = Z_{ik}D_{ik}(1) + (1 - Z_{ik})D_{ik}(0)$.

Step 6. Lastly, we generate a continuous $Y_{ik}(z,d)$ for $z=0,1$ and $d=0,1$, each as a function of $U_{ik}$, $X_{ik}$, and $V_{ik}(z)$. To simplify, the slopes for $U_{ik}$, $X_{ik}$, and $V_{ik}(z)$ are constant across sites. To satisfy the within-site zero covariance assumption (Assumption 2), the slopes for $U_{ik}$, $X_{ik}$, and $V_{ik}(z)$ do not depend on $z$ or $d$.
$$
\begin{aligned}
    Y_{ik}(0,0) &= 80 + g_{0k} + 20(X_{ik}-\bar X_k) + 40(U_{ik}-\bar U_k) + 0.2V_{ik}(0), \\
    Y_{ik}(0,1) &= 95 + g_{0k} + g_{dk} + 20(X_{ik}-\bar X_k) + 40(U_{ik}-\bar U_k) + 0.2V_{ik}(0), \\
    Y_{ik}(1,0) &= 90 + g_{0k} + g_{zk} + 20(X_{ik}-\bar X_k) + 40(U_{ik}-\bar U_k) + 0.2V_{ik}(1), \\
    Y_{ik}(1,1) &= 100 + g_{0k} + g_{zk} + g_{dk} + g_{zdk} + 20(X_{ik}-\bar X_k) + 40(U_{ik}-\bar U_k) + 0.2V_{ik}(1),
\end{aligned}
$$
where $g_{0k}\sim N(0,3^2)$, $g_{zk}\sim N(0,2^2)$, $g_{dk}\sim N(0,2^2)$, and $g_{zdk}\sim N(0,1)$ are mutually independent.

The corresponding key parameters in our theoretical model are as follows:

$$
\begin{aligned}
    Y_{ik}(z,d) = & \theta_{0k} + \theta_{X0k}(X_{ik}-\bar X_k) + \theta_{U0k}(U_{ik}-\bar U_k) + \theta_{Vk}V_{ik}(z) \\
    & + \gamma_{1k}z + \gamma_{2k}d + \gamma_{3k}zd + e_{zd.ik}^{(Y)}.
\end{aligned}
$$
$$
\begin{aligned}
    \theta_{0k} &= 80 + g_{0k},\quad \theta_{X0k} = 20,\quad \theta_{U0k} = 40,\quad \theta_{Vk} = 0.2, \\
    \gamma_{1k} &= 10 + g_{zk},\quad \gamma_{2k} = 15 + g_{dk},\quad \gamma_{3k} = -5 + g_{zdk}.
\end{aligned}
$$
Finally, the observed outcome is $Y_{ik} = Z_{ik}D_{ik}Y_{ik}(1,1) + Z_{ik}(1-D_{ik})Y_{ik}(1,0) + (1-Z_{ik})D_{ik}Y_{ik}(0,1) + (1-Z_{ik})(1-D_{ik})Y_{ik}(0,0) + e^{(Y)}_{ik}$, where $e^{(Y)}_{ik}\sim N(0,6^2)$.

Step 7. Given the parameter values in this data generation procedure, the true cumulative average treatment effect is
$$
ATE = \gamma_1 + \gamma_2 + \gamma_3 + \alpha_1 \theta_V = 10 + 15 - 5 + 5\times0.2 = 21.
$$

In the second set of simulations, in order to examine the consequence of violating the within-site zero covariance assumption (Assumption 2), we make $\gamma_{1ik}$, $\gamma_{2ik}$, and $\gamma_{3ik}$ each be a function of $X_{ik}$ and $U_{ik}$ in the model for $Y_{ik}(z,d)$. Specifically,

$$
\begin{aligned}
    Y_{ik}(0,0) &= 80 + g_{0k} + 20(X_{ik}-\bar X_k) + 40(U_{ik}-\bar U_k) + 0.2V_{ik}(0), \\
    Y_{ik}(0,1) &= 95 + g_{0k} + g_{dk} + 25(X_{ik}-\bar X_k) + 40(U_{ik}-\bar U_k) + 0.2V_{ik}(0), \\
    Y_{ik}(1,0) &= 90 + g_{0k} + g_{zk} + 20(X_{ik}-\bar X_k) + 45(U_{ik}-\bar U_k) + 0.2V_{ik}(1), \\
    Y_{ik}(1,1) &= 100 + g_{0k} + g_{zk} + g_{dk} + g_{zdk} + 25(X_{ik}-\bar X_k) + 50(U_{ik}-\bar U_k) + 0.2V_{ik}(1),
\end{aligned}
$$
where $g_{0k}\sim N(0,3^2)$, $g_{zk}\sim N(0,2^2)$, $g_{dk}\sim N(0,2^2)$, and $g_{zdk}\sim N(0,1)$ are mutually independent.

The corresponding key parameters in our theoretical model are as follows:

$$
\begin{aligned}
    Y_{ik}(z,d) = & \theta_{0k} + \theta_{X0k}(X_{ik}-\bar X_k) + \theta_{U0k}(U_{ik}-\bar U_k) + \theta_{Vk}V_{ik}(z) \\
    & + \gamma_{1k}z + \gamma_{2k}d + \gamma_{3k}zd + e_{zd.ik}^{(Y)}.
\end{aligned}
$$
$$
\begin{aligned}
    \theta_{0k} &= 80 + g_{0k},\quad \theta_{X0k} = 20,\quad \theta_{U0k} = 40,\quad \theta_{Vk} = 0.2, \\
    \gamma_{1ik} &= 10 + g_{zk} + 5(U_{ik} - \bar{U}_{ik}),\quad \gamma_{2ik} = 15 + g_{dk} + 5(X_{ik} - \bar{X}_{ik}), \\
    \gamma_{3ik} &= -5 + g_{zdk} + 5(U_{ik} - \bar{U}_{ik}).
\end{aligned}
$$
Finally, the observed outcome is $Y_{ik} = Z_{ik}D_{ik}Y_{ik}(1,1) + Z_{ik}(1-D_{ik})Y_{ik}(1,0) + (1-Z_{ik})D_{ik}Y_{ik}(0,1) + (1-Z_{ik})(1-D_{ik})Y_{ik}(0,0) + e^{(Y)}_{ik}$, where $e^{(Y)}_{ik}\sim N(0,6^2)$.

Note that $\beta_{1ik}$ and $\beta_{2ik}$ are already each a function of $X_{ik}$ and $U_{ik}$ in the model for $D_{ik}(z)$. Therefore, $\mathrm{Cov}(\gamma_{2ik}, \beta_{1ik}) \ne 0$, $\mathrm{Cov}(\gamma_{3ik}, \beta_{2ik}) \ne 0$, and the within-site zero covariance assumption (Assumption 2) is violated.

In the third set of simulations, we examine the consequence of violating the between-site independence assumption (Assumption 3) by replacing $g_{zk}$ in the $Y$ model by $\frac{1}{3}t_{1k}$, so that $\alpha_{1k} = 5 + t_{1k}$ and $\gamma_{1k} = 10 + \frac{1}{3}t_{1k}$ are correlated.

\section{Comparisons Between Analytic Methods}

We compare the estimation performance of our proposed MS2T-IV strategy with that of the alternative analytic methods:

\vspace{-0.2cm}
(a) Naïve analysis without adjustment for $X$ or $V$.
$$
ATE_k = E(Y|Z=1,D=1,k) - E(Y|Z=0,D=0,k),
$$
and $ATE = E(ATE_k)$.

(b) Covariance adjustment for $X$ via linear regression.
$$
ATE_k = E_x\left\{E(Y|Z=1,D=1,X=x,k) - E(Y|Z=0,D=0,X=x,k)\right\}
$$
and $ATE = E(ATE_k)$.

(c) IPTW analysis that adjusts for $X$ and $V$. In each site, we estimate $Pr(D=d \mid Z=z,X=x,V=v,k)$ and $Pr(D=d \mid Z=z,k)$. For individual $i$ in site $k$ with $D_{ik}=d, Z_{ik}=z, V_{ik}=v$, the weight is
$$
w_{ik} = \frac{Pr(D=d \mid Z=z,k)}{Pr(D=d \mid Z=z,X=x,V=v,k)}.
$$
Then $ATE_k = E[wY \mid Z=1,D=1,k] - E[wY \mid Z=0,D=0,k]$, and $ATE = E[ATE_k]$.

(d) Our proposed MS2T-IV method without adjustment for $X$ in Stage 1 analysis.

(e) Our proposed MS2T-IV method with adjustment for $X$ in Stage 1 analysis.

The evaluation criteria for estimation performance include bias, efficiency, and mean squared error (MSE). The bias in the ATE estimator is computed as $\overline{\hat\delta_{ATE}}-\delta_{ATE}$, where $\overline{\hat\delta_{ATE}}$ is the average value of the ATE estimate $\hat\delta_{ATE}$ over 500 simulations. The efficiency of the ATE estimator is assessed using the empirical variance of $\hat\delta_{ATE}$ from 500 simulations, and the MSE is computed as the sum of the squared bias and the empirical variance.

\section{Estimation Performance When the Assumptions are Violated}

\begin{table}[htbp]
\centering
\caption{Estimation Performance (Assumption 2 Violated)}
\label{table: A2 violated}
\begin{tabular}{@{}lcrrrrrrrr@{}}
\toprule
&& \multicolumn{4}{c}{$K=25$} & \multicolumn{4}{c}{$K=100$} \\
\cmidrule(lr){3-6}\cmidrule(lr){7-10}
& $n_k$ & 30 & 100 & 1{,}000 & 5{,}000 & 30 & 100 & 1{,}000 & 5{,}000 \\
\midrule
\multicolumn{9}{@{}l}{\textbf{Bias}} \\
Naive & & 2.75 & 2.71 & 2.75 & 2.76 & 2.69 & 2.70 & 2.75 & 2.71 \\
Naive-adj & & 2.05 & 2.04 & 2.10 & 2.11 & 1.93 & 2.06 & 2.11 & 2.06 \\
IPTW & & -0.99 & -0.31 & 0.51 & 0.65 & -1.16 & -0.30 & 0.45 & 0.53 \\
MS2T-IV & & 0.04 & 0.09 & 0.17 & 0.12 & 0.05 & 0.11 & 0.18 & 0.16 \\
MS2T-IV-adj & & -0.01 & 0.05 & 0.15 & 0.12 & 0.01 & 0.05 & 0.16 & 0.15 \\
\addlinespace[0.6ex]
\multicolumn{9}{@{}l}{\textbf{Empirical Variance}} \\
Naive & & 7.09 & 3.50 & 1.18 & 1.09 & 1.92 & 0.74 & 0.31 & 0.28 \\
Naive-adj & & 6.46 & 3.02 & 0.91 & 0.87 & 1.88 & 0.65 & 0.24 & 0.22 \\
IPTW & & 29.67 & 9.94 & 5.04 & 4.42 & 7.86 & 2.43 & 1.30 & 1.26 \\
MS2T-IV & & 11.43 & 5.27 & 1.82 & 1.68 & 2.65 & 1.40 & 0.47 & 0.32 \\
MS2T-IV-adj & & 10.47 & 4.37 & 1.73 & 1.64 & 2.47 & 1.22 & 0.44 & 0.31 \\
\addlinespace[0.6ex]
\multicolumn{9}{@{}l}{\textbf{MSE}} \\
Naive & & 14.65 & 10.82 & 8.76 & 8.72 & 9.16 & 8.03 & 7.89 & 7.63 \\
Naive-adj & & 10.68 & 7.17 & 5.33 & 5.32 & 5.62 & 4.90 & 4.69 & 4.47 \\
IPTW & & 30.65 & 10.04 & 5.30 & 4.84 & 9.20 & 2.52 & 1.50 & 1.54 \\
MS2T-IV & & 11.44 & 5.28 & 1.85 & 1.70 & 2.65 & 1.41 & 0.50 & 0.35 \\
MS2T-IV-adj & & 10.47 & 4.37 & 1.75 & 1.66 & 2.47 & 1.23 & 0.46 & 0.33 \\
\bottomrule
\end{tabular}
\end{table}

\begin{table}[htbp]
\centering
\caption{Estimation Performance (Assumption 3 Violated)}
\label{table: A3 violated}
\begin{tabular}{@{}lcrrrrrrrr@{}}
\toprule
&& \multicolumn{4}{c}{$K=25$} & \multicolumn{4}{c}{$K=100$} \\
\cmidrule(lr){3-6}\cmidrule(lr){7-10}
& $n_k$ & 30 & 100 & 1{,}000 & 5{,}000 & 30 & 100 & 1{,}000 & 5{,}000 \\
\midrule
\multicolumn{9}{@{}l}{\textbf{Bias}} \\
Naive & & 1.80 & 1.73 & 1.79 & 1.83 & 1.75 & 1.74 & 1.78 & 1.74 \\
Naive-adj & & 1.36 & 1.30 & 1.37 & 1.41 & 1.27 & 1.33 & 1.37 & 1.33 \\
IPTW & & -2.62 & -1.52 & -0.51 & -0.34 & -2.83 & -1.50 & -0.59 & -0.50 \\
MS2T-IV & & -0.01 & 0.00 & 0.03 & 0.03 & -0.02 & -0.04 & 0.03 & 0.00 \\
MS2T-IV-adj & & -0.03 & -0.01 & 0.01 & 0.03 & -0.04 & -0.09 & 0.01 & 0.00 \\
\addlinespace[0.6ex]
\multicolumn{9}{@{}l}{\textbf{Empirical Variance}} \\
Naive & & 5.34 & 2.67 & 1.06 & 0.99 & 1.45 & 0.58 & 0.26 & 0.26 \\
Naive-adj & & 4.92 & 2.40 & 0.88 & 0.86 & 1.41 & 0.52 & 0.22 & 0.22 \\
IPTW & & 25.33 & 8.82 & 4.80 & 4.17 & 6.67 & 2.15 & 1.22 & 1.21 \\
MS2T-IV & & 7.52 & 3.40 & 1.26 & 1.23 & 1.81 & 0.82 & 0.30 & 0.24 \\
MS2T-IV-adj & & 6.80 & 2.99 & 1.18 & 1.21 & 1.67 & 0.73 & 0.28 & 0.23 \\
\addlinespace[0.6ex]
\multicolumn{9}{@{}l}{\textbf{MSE}} \\
Naive & & 8.57 & 5.68 & 4.25 & 4.33 & 4.50 & 3.60 & 3.42 & 3.29 \\
Naive-adj & & 6.78 & 4.10 & 2.75 & 2.84 & 3.02 & 2.30 & 2.08 & 1.97 \\
IPTW & & 32.20 & 11.12 & 5.06 & 4.28 & 14.65 & 4.41 & 1.57 & 1.46 \\
MS2T-IV & & 7.52 & 3.40 & 1.27 & 1.23 & 1.81 & 0.83 & 0.30 & 0.24 \\
MS2T-IV-adj & & 6.81 & 2.99 & 1.18 & 1.21 & 1.67 & 0.74 & 0.28 & 0.23 \\
\bottomrule
\end{tabular}
\end{table}

\end{document}